\documentclass[iop]{emulateapj}
\usepackage{enumerate,multirow,amsmath}
\shorttitle{The M$_\mathrm{BH}$ -- M$_\mathrm{sph,*}$ relation.}
\shortauthors{Scott, Graham \& Schombert}

\begin{document}

\title{The supermassive black hole mass -- spheroid stellar mass relation for S\'ersic and core-S\'ersic galaxies}
\author{Nicholas Scott, Alister W Graham}
\affil{Centre for Astrophysics and Supercomputing, Swinburne University of Technology, Hawthorn, Vic, 3122, Australia}
\author{and \\ James Schombert}
\affil{Department of Physics, University of Oregon, Eugene, OR 97403, USA}

\begin{abstract}
We have examined the relationship between supermassive black hole mass (M$_\mathrm{BH}$) and the stellar mass of the host spheroid (M$_\mathrm{sph,*}$) for a sample of 75 nearby galaxies. To derive the spheroid stellar masses we used improved 2MASS {\it K$_\mathrm{\it s}$}-band photometry from the {\sc archangel} photometry pipeline. Dividing our sample into core-S\'ersic and S\'ersic galaxies, we find that they are described by very different M$_\mathrm{BH}$--M$_\mathrm{sph,*}$ relations. For core-S\'ersic galaxies --- which are typically massive and luminous, with M$_{\rm BH} \gtrsim 2\times10^8 {\rm M}_\odot$ --- we find M$_\mathrm{BH} \propto$ M$_\mathrm{sph,*}^{0.97 \pm 0.14}$, consistent with other literature relations. However, for the S\'ersic galaxies --- with typically lower masses, M$_\mathrm{sph,*} \lesssim 3\times10^{10}\ \rm{M}_\odot$ --- we find M$_\mathrm{BH} \propto \rm{M}_\mathrm{sph,*}^{2.22 \pm 0.58}$, a dramatically steeper slope that differs by more than 2 standard deviations. This relation confirms that, for S\'ersic galaxies, M$_\mathrm{BH}$ {\it is not} a constant fraction of M$_\mathrm{sph,*}$. S\'ersic galaxies can grow via the accretion of gas which fuels both star formation and the central black hole, as well as through merging. Their black hole grows significantly more rapidly than their host spheroid, prior to growth by dry merging events that produce core-S\'ersic galaxies, where the black hole and spheroid grow in lock step. We have additionally compared our S\'ersic M$_\mathrm{BH}$--M$_\mathrm{sph,*}$ relation with the corresponding relation for nuclear star clusters, confirming that the two classes of central massive object follow significantly different scaling relations.
\end{abstract}


\section{Introduction}
\label{sec:intro}

Supermassive black hole masses, M$_\mathrm{BH}$, are well known to scale with a number of properties of their host galaxy. This was first reported by \citet{Kormendy:1995}, who found a linear correlation between supermassive black hole mass and host spheroid luminosity. Later studies found log-linear correlations between supermassive black hole mass and: stellar velocity dispersion, $\sigma$ \citep{Ferrarese:2000,Gebhardt:2000}, stellar concentration \citep{Graham:2001} and dynamical mass, $M_\mathrm{dyn} \propto \sigma^2 R$ \citep{Magorrian:1998,Marconi:2003,Haring:2004}. These initial studies reported strong log-linear correlations.

Recent work using larger galaxy samples, with accurate measurements of their black hole masses and host galaxy properties, has indicated that these simple log-linear scaling relations are not always sufficient descriptions of the observed distribution. In particular, \citet{Graham:2007b} showed that the black hole mass - S\'ersic index ($n$) relation was not log-linear, and \citet{Graham:2012a} showed that the M$_\mathrm{BH}$--M$_\mathrm{sph,dyn}$ relation requires two separate log-linear relations: with a slope of $\sim 2$ for S\'ersic galaxies \citep[whose bulge surface brightness profiles are well-represented by a single][model\footnote{The S\'ersic model is reviewed in detail in \citet{Graham:2005}}]{Sersic:1968} and with a slope $\sim 1$ for core-S\'ersic galaxies \citep[whose bulge surface brightness profiles deviate from a single S\'ersic function by having a partially depleted core:][]{Graham:2003a,Graham:2003,Trujillo:2004,Ferrarese:2006b}. In addition \citet[hereafter GS13]{Graham:2013} demonstrated that the M$_\mathrm{BH}$--L$_\mathrm{sph}$ relation is also better described by two log-linear relations, with {\it K$_\mathrm{s}$}-band slopes $\sim 2.73\pm0.55$ and $\sim 1.10\pm0.20$ for the S\'ersic and core-S\'ersic galaxies respectively. In contrast the M$_\mathrm{BH}$--$\sigma$ relation {\it is} well described by a single log-linear relation with slope $\sim 5$ \citep[][GS13]{Ferrarese:2000,Graham:2011,McConnell:2013,Park:2012} over this same black hole mass range, once the offset barred and pseudobulge galaxies are properly taken into account \citep{Graham:2008,Hu:2008,Graham:2009b}. 

These revisions to the supermassive black hole scaling relations are in fact expected, given other observed galaxy scaling relations. As observational samples have explored a broader range in galaxy luminosity, the L--$\sigma$ relation has been revised from L $\propto \sigma^4$ \citep{Faber:1976} to exhibit two different slopes at high and low luminosity.  The relation for luminous galaxies is given by L $\propto \sigma^5$ \citep{Schechter:1980,Liu:2008} but for M$_B > -20.5$ mag L $\propto \sigma^2$ \citep{Davies:1983,Matkovic:2005}. Given this form of the L--$\sigma$ relation, and the log-linear M$_\mathrm{BH}$--$\sigma$ relation, the M$_\mathrm{BH}$--L relation {\it cannot} be log-linear --- it must exhibit the same bend as the L--$\sigma$ relation. From these correlations, the expected form of the M$_\mathrm{BH}$--L$_\mathrm{sph}$ relation is: M$_\mathrm{BH} \propto$ L$_\mathrm{sph}^{1.0}$ for luminous core-S\'ersic galaxies and M$_\mathrm{BH} \propto$ L$_\mathrm{sph}^{2.5}$ for the typically less-luminous S\'ersic galaxies, consistent with the findings of GS13.

\begin{figure}[t]
\includegraphics[width=3.25in]{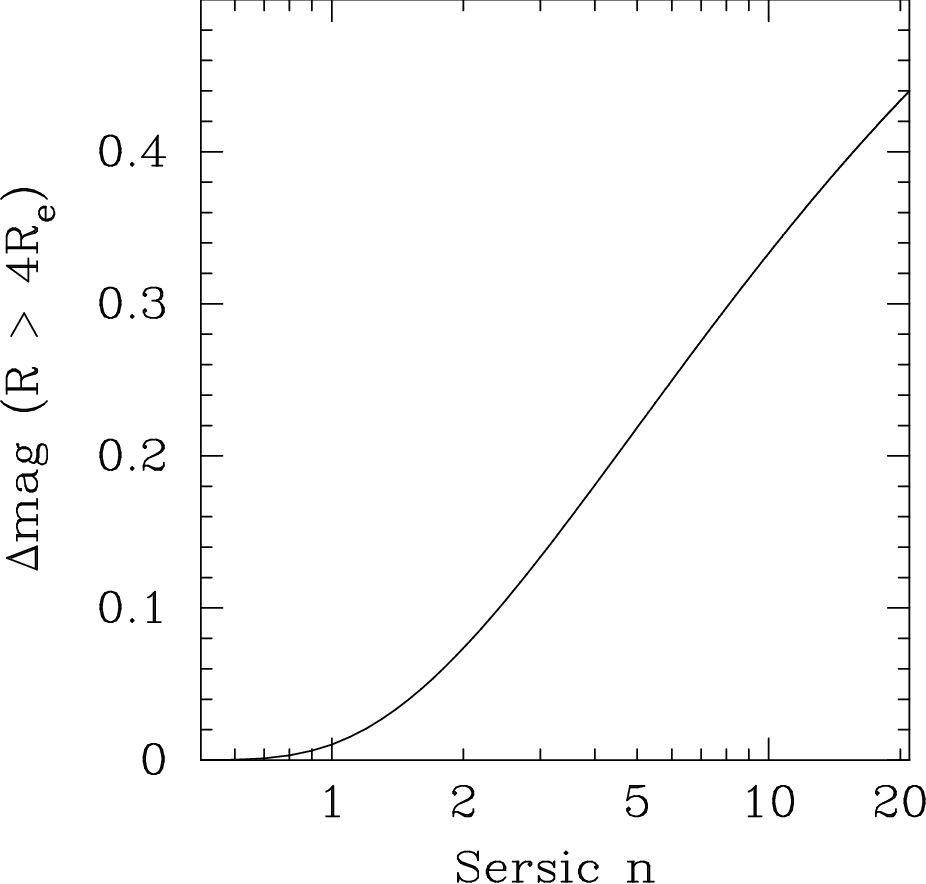}
\caption{The total flux found at radii greater than four effective radii for S\'ersic surface brightness profiles as a function of S\'ersic index $n$. For galaxies with $n \gtrsim 3$, the flux found at large radii is significant. Total magnitudes in the 2MASS Extended Source Catalogue miss this additional flux at large radius, however this is not the case for our {\sc archangel} derived total magnitudes.}
\label{fig:flux_at_r}
\end{figure}

The steep M$_{\rm BH}$--L$_{\rm sph}$ relation for S\'ersic galaxies implies that their black hole must grow much more rapidly than their host spheroid. In these intermediate-mass galaxies, the accretion of gas plays a significant role in the growth of the spheroid and in fuelling the central supermassive black hole. This is seen at high redshift through the coexistence of Active Galactic Nuclei (AGN) with rapidly star-forming sub-millimetre galaxies \citep[e.g.,][]{Blain:1999,Page:2001,Alexander:2005,Pope:2008,Page:2012,Simpson:2012} and ultra-luminous FIR-detected galaxies \citep[e.g.,][]{Norman:1988,Alonso-Herrero:2006,Daddi:2007,Murphy:2009}, and through large spectroscopic surveys of AGN hosts \citep[e.g.,][]{Silverman:2009,Chen:2013}. At low redshift this is evident from the coincidence of ongoing star formation or young stellar populations and AGN activity \citep[e.g.,][]{Kauffmann:2003b,Netzer:2009,Rosario:2013}. These observations are all consistent with a model in which there is a strong physical link between star formation and AGN activity \citep{Hopkins:2006}. In a large sample of low-redshift AGN, \citet{LaMassa:2013} find that the star formation rate is related to the black hole accretion rate by: $\dot{M} \propto SFR^{2.78}$, indicating that gas accretion contributes much more significantly to black hole growth than to star formation.

An accurate determination of the true form of the supermassive black hole scaling relations is critical in a number of areas of extragalactic astrophysics. Supermassive black holes play a critical role in semi-analytic models of galaxy formation, through their ability to regulate star formation via AGN feedback \citep{SIlk:1998,Kauffmann:2000}. This feedback is vital in matching the predicted galaxy/bulge luminosity functions of such models to observations. Modern semi-analytic models use the observed local supermassive black hole scaling relations as a key constraint on the rate of black hole growth \citep[e.g.,][]{Springel:2005,Bower:2006,Croton:2006,Di-Matteo:2008,Booth:2009,Dubois:2012}. Using an incorrect form of the local scaling relations can significantly alter the degree of AGN feedback in these simulations, resulting in an inaccurate determination of the efficacy of that feedback, or in incorrect rates of star formation and build-up of stellar mass.

Another important application of the local supermassive black hole scaling relations is to the study of the evolution of the black hole -- host spheroid connection with redshift. The M$_\mathrm{BH}$--L$_\mathrm{sph}$ and M$_\mathrm{BH}$--M$_\mathrm{sph,*}$ relations have been determined over a range of redshifts, including at $z\lesssim0.5$ \citep{Woo:2006,Salviander:2007,Shen:2008,Canalizo:2012,Hiner:2012}, $z\sim1$ \citep{Peng:2006,Salviander:2007,Merloni:2010,Bennert:2011,Bluck:2011,Zhang:2012} and $z>2$ \citep{Borys:2005,Kuhlbrodt:2005,Peng:2006,Shields:2006}. These studies typically search for changes in the high-redshift scaling relations with respect to the local scaling relations in an effort to identify evolution in the relationship between supermassive black holes and their host spheroids. From any apparent evolution, they then attempt to determine whether the onset of spheroid or black hole growth occurred first, and therefore which is the driving mechanism for the local correlations. Correctly determining the local scaling relations is therefore critical in identifying any evolution with redshift --- as the local scaling relations are much more accurately known than those at high redshift much of the constraint on evolution comes from the local scaling relations. Using an incorrect local scaling relation can hide (or incorrectly identify) evolution with redshift in the black hole -- host spheroid connection.

In this work we extend the investigation of the bent supermassive black hole scaling relations to include the stellar mass of the host spheroid; the M$_\mathrm{BH}$--M$_\mathrm{sph,*}$ relation. In Section \ref{sec:sample} we present our sample of galaxies containing supermassive black holes and describe the derivation of their associated spheroid's stellar luminosity and mass. The luminosities used here differ slightly from those in GS13 due to our use of updated photometry, which we describe below. In Section \ref{sec:analysis} we use a linear regression to examine the bent nature of the M$_\mathrm{BH}$--M$_\mathrm{sph,*}$ relation. In Section \ref{sec:discussion} we discuss the implications of bent supermassive black hole scaling relations on the intrinsic scatter of supermassive black hole scaling relations, the coevolution of supermassive black holes and their host spheroids and the alleged common origins of nuclear star clusters and supermassive black holes and . We present our conclusions in Section \ref{sec:conclusions}.

\section{Sample and Data}
\label{sec:sample}

We make use of the sample of 78 supermassive black hole masses and host spheroid magnitudes presented in GS13. Following GS13, we continue to exclude M32 (due to an unknown amount of stellar stripping), the Milky Way (due to its uncertain bulge magnitude due to dust extinction) and NGC 1316 (due to its uncertain core type), giving a final sample of 75 galaxies. All galaxies in the sample have directly measured supermassive black hole masses, with typical uncertainties $\sim50$\% (for detailed references see GS13, their section 2). GS13 provide apparent {\it B-} and {\it K$_\mathrm{\it s}$-}band magnitude data and a prescription to determine absolute spheroid magnitudes. Although we make use of their {\it B-}band magnitudes, we use different initial {\it K$_\mathrm{\it s}$}-band apparent total galaxy magnitudes as described below.

\begin{figure}[t]
\includegraphics[width=3.25in]{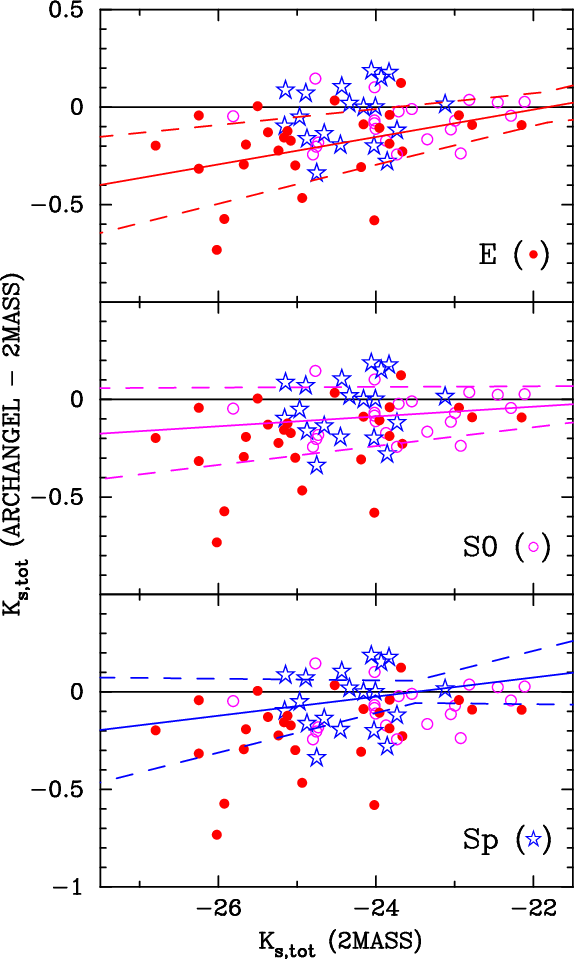}
\caption{Comparison of galaxy total magnitudes between the 2MASS catalogue values and our own {\sc archangel} magnitudes for ellipticals (red filled circles), S0s (pink open circles) and spirals (blue stars). 67/68 galaxies for which we derived {\sc archangel} magnitudes are shown (NGC 2974 was excluded due to a nearby bright star affecting its 2MASS magnitude). Each panel shows the difference between the 2MASS and {\sc archangel} magnitudes as a function of 2MASS magnitude. In each of the upper, middle and lower panels we show linear fits to the data for the elliptical, S0 and spiral subsets of our sample. The solid line shows the best fit, with the dashed lines indicating the $1 \sigma$ uncertainty on each fit. }
\label{fig:photometry_comp}
\end{figure}

GS13 used {\it K$_\mathrm{\it s}$}-band magnitudes from the Two Micron All-Sky Survey (2MASS) Extended Source Catalogue \citep{Jarrett:2000}, however they noted that some errors have recently been reported for the 2MASS catalogue photometry \citep{Schombert:2012}. Given these concerns we derive new total galaxy magnitudes from the same 2MASS {\it K$_\mathrm{\it s}$}-band images using the {\sc archangel} photometry pipeline \citep{Schombert:2012}. The pipeline takes a galaxy's name as input, parsing the name through the NASA Extragalactic Database (NED) to resolve its position on the sky. The pipeline then extracts the four sky strip images surrounding the galaxy's coordinates from the 2MASS Atlas Image Server, stitching the images together to form a single raw image. The 2MASS project provides calibrated, flattened, kernel-smoothed, sky-subtracted images so these steps are not duplicated by the pipeline. An accurate sky determination is then made using user-defined sky boxes clear of obvious stellar or galaxy sources, which are then summed and averaged. The final step is the extraction of isophotal values as a function of radius using an ellipse fitting routine. Elliptical apertures are used to determine curves of growth for determination of photometric values. 

Our new {\it K$_\mathrm{\it s}$} magnitudes improve on the 2MASS Extended Source Catalogue values in three ways. First, as described in \citet{Schombert:2012}, the sky background is over-subtracted by the 2MASS data reduction pipeline, resulting in an underestimation of the total galaxy magnitude. This over-subtraction truncates the surface brightness profile of luminous elliptical galaxies at large radii, causing the 2MASS pipeline to underestimate the physical size of each object. This results in the total magnitude being measured from an aperture that significantly underestimates the true physical size of an object, thus resulting in a total magnitude that is 10--40 \% lower than the total magnitudes derived by {\sc archangel}. Finally, our {\sc archangel} derived magnitudes are determined from pipeline fits to the surface brightness profiles, extending out to radii where the uncertainty in the surface brightness of the object exceeds 1 mag arcsec$^{-2}$, which is typically further than the four-half-light-radii apertures used for 2MASS total magnitudes. For low S\'ersic index profiles there is little difference in the total magnitudes derived from these two apertures, however, as shown in Figure \ref{fig:flux_at_r}, for S\'ersic $n \gtrsim 3$ the flux missed by the 2MASS aperture can be significant.

In Figure \ref{fig:photometry_comp} we show a comparison between the 2MASS catalogue magnitudes and our new {\sc archangel} magnitudes for 67 galaxies. Eight galaxies of our 75 were too large on the sky to model readily with {\sc archangel}. For these eight remaining galaxies we derived a correction to their 2MASS magnitudes using the following procedure. We divided the sample of 67 galaxies into three morphological types (ellipticals, lenticulars and spirals) and fit linear relations to the 2MASS vs.\ {\sc archangel} magnitudes for each subset (indicated by the lines in Figure \ref{fig:photometry_comp}). Based on these relations and the 2MASS catalogue magnitude we derived corrected magnitudes for the 2/8 remaining elliptical galaxies and used the 2MASS magnitudes for the 6/8 S0 and Sa disk galaxies. The relation between the 2MASS and {\sc archangel} {\it K$_\mathrm{\it s}$}-band magnitudes for elliptical galaxies in our sample is:
\begin{equation}
K_{s,\mathrm{ARCH}} = (1.070 \pm 0.030) K_{s,\mathrm{2MASS}} + (1.527 \pm 0.061).
\end{equation}
The two galaxies for which this procedure was used are indicated with a $\dagger$ in Table \ref{tab:sample}, along with the six disk galaxies for which we used 2MASS photometry. As tabulated in GS13, total apparent {\it B-}band magnitudes were drawn from the Third Reference Catalogue of Bright Galaxies \citep{deVaucouleurs:1991}. 

Following GS13, apparent magnitudes were converted to absolute magnitudes using distance moduli primarily from the surface brightness fluctuation based measurements of \citet{Tonry:2001}, after applying the 0.06 magnitude correction of \citet[][see GS13 for a full list of references for the distance determinations]{Blakeslee:2002}. All magnitudes were corrected for Galactic extinction following \citet{Schlegel:1998}, cosmological redshift dimming and {\it K}-corrections.

\begin{table}[t]
\caption{Properties of our sample of 75 nearby galaxies hosting a supermassive black hole}
\label{tab:sample}
\begin{center}
\begin{tabular}{l c c c c c c}
\hline
Galaxy & Type & Core & M$_\mathrm{BH}$ & $K_\mathrm{s,sph}$ & $B-K_s$ & M$_\mathrm{sph,*}$ \\
& & & [$10^8$ M$_\odot$] & [mag] & [mag] & [$10^{10}$ M$_\odot$] \\
\hline
Abel1836 & BCG & y? & 39$^{+    4}_{-    5}$ & -26.29 & 4.57 & 69$^{+59}_{-32}$ \\
Abel3565 & BCG & y? & 11$^{+    2}_{-    2}$ & -25.84 & 4.13 & 37$^{+32}_{-17}$ \\
CenA$^{\dagger}$ & S0 & n? & 0.45$^{+ 0.17}_{- 0.10}$ & -22.87 & 3.03 & 1.4$^{+2.0}_{-0.8}$ \\
CygnusA$^{\dagger}$ & S? & y? & 25$^{+    7}_{-    7}$ & -25.81 & 5.01 & 55$^{+80}_{-33}$ \\
IC1459 & E3 & y? & 24$^{+   10}_{-   10}$ & -25.50 & 4.13 & 27$^{+23}_{-12}$ \\
IC2560 & SBb & n? & 0.044$^{+0.044}_{-0.022}$ & -22.97 & 3.95 & 2.4$^{+3.5}_{-1.4}$ \\
M31$^{\dagger}$ & Sb & n & 1.4$^{+  0.9}_{-  0.3}$ & -21.69 & 3.02 & 0.46$^{+0.68}_{-0.28}$ \\
M81$^{\dagger}$ & Sab & n & 0.73$^{+??}_{-??}$ & -22.56 & 2.97 & 1.0$^{+1.5}_{-0.6}$ \\
M84 & E1 & y &  9.0$^{+  0.9}_{-  0.8}$ & -25.25 & 3.90 & 19$^{+16}_{-9}$ \\
M87$^{\dagger}$ & E0 & y & 58$^{+  3.5}_{-  3.5}$ & -25.43 & 3.96 & 23$^{+  19}_{-  10}$ \\
NGC253$^{\dagger}$ & SBc & n & 0.10$^{+ 0.10}_{- 0.05}$ & -21.42 & 4.09 & 0.61$^{+0.89}_{-0.36}$ \\
NGC524 & S0 & y & 8.3$^{+  2.7}_{-  1.3}$ & -23.82 & 3.67 & 4.6$^{+6.6}_{-2.7}$ \\
NGC821 & E & n & 0.39$^{+ 0.26}_{- 0.09}$ & -24.60 & 3.95 & 11$^{+9}_{-5}$ \\
NGC1023 & SB0 & n & 0.42$^{+ 0.04}_{- 0.04}$ & -23.07 & 3.25 & 1.9$^{+2.7}_{-1.1}$ \\
NGC1068 & SBb & n & 0.084$^{+0.003}_{-0.003}$ & -23.57 & 4.12 & 4.5$^{+6.6}_{-2.7}$ \\
NGC1194 & S0 & n? & 0.66$^{+ 0.03}_{- 0.03}$ & -22.80 & 3.22 &1.4$^{+2.1}_{-0.8}$ \\
NGC1300 & SBbc & n & 0.73$^{+ 0.69}_{- 0.35}$ & -21.73 & 3.69 &0.66$^{+0.97}_{-0.40}$ \\
NGC1332 & S0 & y? & 15$^{+    2}_{-    2}$ & -23.92 & 3.55 & 4.7$^{+6.9}_{-2.8}$ \\
NGC1399 & E & y & 4.7$^{+  0.6}_{-  0.6}$ & -25.32 & 4.37 & 26$^{+22}_{-12}$ \\
NGC2273 & SBa & n & 0.083$^{+0.004}_{-0.004}$ & -22.94 & 3.61 & 2.0$^{+2.9}_{-1.2}$ \\
NGC2549 & SB0 & n & 0.14$^{+ 0.02}_{- 0.13}$ & -21.42 & 3.18 & 0.39$^{+0.57}_{-0.23}$ \\
NGC2778 & SB0 & n & 0.15$^{+ 0.09}_{-  0.1}$ & -21.19 & 3.40 & 0.35$^{+0.52}_{-0.21}$ \\
NGC2787 & SB0 & n & 0.40$^{+ 0.04}_{- 0.05}$ & -20.96 & 3.55 & 0.30$^{+0.45}_{-0.18}$ \\
NGC2960 & Sa? & n? & 0.12$^{+0.005}_{-0.005}$ & -23.85 & 3.06 & 3.5$^{+5.1}_{-2.1}$ \\
NGC2974 & E & n &  1.7$^{+  0.2}_{-  0.2}$ & -24.03 & 4.06 & 6.7$^{+5.7}_{-3.1}$ \\
NGC3079 & SBc & n & 0.024$^{+0.024}_{-0.012}$ & -21.84 & 4.04 & 0.88$^{+1.28}_{-0.52}$ \\
NGC3115 & S0 & n & 8.8$^{+ 10.0}_{-  2.7}$ & -23.09 & 3.21 & 1.9$^{+2.7}_{-1.1}$ \\
NGC3227 & SBa & n & 0.14$^{+ 0.10}_{- 0.06}$ & -22.60 & 2.73 & 0.93$^{+1.37}_{-0.56}$ \\
NGC3245 & S0 & n &  2.0$^{+  0.5}_{-  0.5}$ & -22.63 & 3.26 & 1.24$^{+1.8}_{-0.7}$ \\
NGC3368 & SBab & n & 0.073$^{+0.015}_{-0.015}$ & -22.29 & 3.15 & 0.86$^{+1.26}_{-0.51}$ \\
NGC3377 & E5 & n & 0.77$^{+ 0.04}_{- 0.06}$ & -22.87 & 3.78 & 2.0$^{+1.7}_{-0.9}$ \\
NGC3379 & E1 & y &  4.0$^{+  1.0}_{-  1.0}$ & -23.87 & 3.94 & 5.4$^{+4.7}_{-2.5}$ \\
NGC3384 & SB0 & n & 0.17$^{+ 0.01}_{- 0.02}$ & -22.48 & 3.43 & 1.2$^{+1.7}_{-0.7}$ \\
NGC3393 & SBa & n & 0.34$^{+ 0.02}_{- 0.02}$ & -23.89 & 3.68 & 4.9$^{+7.1}_{-2.9}$ \\
NGC3414 & S0 & n & 2.4$^{+  0.3}_{-  0.3}$ & -22.98 & 3.58 & 2.0$^{+2.9}_{-1.2}$ \\
NGC3489 & SB0 & n & 0.058$^{+0.008}_{-0.008}$ & -21.95 & 3.26 & 0.66$^{+0.97}_{-0.40}$ \\
NGC3585 & S0 & n & 3.1$^{+  1.4}_{-  0.6}$ & -23.94 & 3.67 & 5.1$^{+7.4}_{-3.0}$ \\
NGC3607 & S0 & n &  1.3$^{+  0.5}_{-  0.5}$ & -23.53 & 3.06 & 2.6$^{+3.8}_{-1.5}$ \\
NGC3608 & E2 & y &  2.0$^{+  1.1}_{-  0.6}$ & -23.56 & 3.43 & 3.2$^{+2.7}_{-1.5}$ \\
NGC3842 & E & y & 97$^{+   30}_{-   26}$ & -26.75 & 4.47 & 100$^{+86}_{-46}$ \\
NGC3998 & S0 & y? & 8.1$^{+  2.0}_{-  1.9}$ & -22.37 & 3.96 & 1.4$^{+2.1}_{-0.9}$ \\
NGC4026 & S0 & n &  1.8$^{+  0.6}_{-  0.3}$ & -22.22 & 3.34 & 0.88$^{+1.30}_{-0.53}$ \\
NGC4151 & SBab & n & 0.65$^{+ 0.07}_{- 0.07}$ & -22.60 & 3.36 & 1.3$^{+1.9}_{-0.8}$ \\
NGC4258 & SBbc & n & 0.39$^{+ 0.01}_{- 0.01}$ & -21.58 & 3.63 & 0.56$^{+0.82}_{-0.34}$ \\
NGC4261 & E2 & y &  5.0$^{+  1.0}_{-  1.0}$ & -25.46 & 4.35 & 29$^{+25}_{-13}$ \\
NGC4291 & E2 & y & 3.3$^{+  0.9}_{-  2.5}$ & -23.89 & 4.13 & 6.1$^{+5.2}_{-2.8}$ \\
NGC4342 & S0 & n &  4.5$^{+  2.3}_{-  1.5}$ & -21.70 & 3.71 & 0.65$^{+0.96}_{-0.39}$ \\
NGC4388 & Sb & n? & 0.075$^{+0.002}_{-0.002}$ & -23.57 & 3.50 & 3.3$^{+4.9}_{-2.0}$ \\
NGC4459 & S0 & n & 0.68$^{+ 0.13}_{- 0.13}$ & -22.90 & 3.75 & 2.0$^{+3.0}_{-1.2}$ \\
NGC4473 & E5 & n &  1.2$^{+  0.4}_{-  0.9}$ & -24.02 & 4.13 & 6.9$^{+5.9}_{-3.2}$ \\
NGC4486a & E2 & n & 0.13$^{+ 0.08}_{- 0.08}$ & -22.24 & 4.19 & 1.4$^{+1.2}_{-0.6}$ \\
NGC4552 & E & y & 4.7$^{+  0.5}_{-  0.5}$ & -24.25 & 3.94 & 7.7$^{+6.6}_{-3.6}$ \\
NGC4564 & S0 & n & 0.59$^{+ 0.03}_{- 0.09}$ & -22.10 & 3.61 & 0.90$^{+1.3}_{-0.5}$ \\
NGC4594 & Sa & y & 6.4$^{+  0.4}_{-  0.4}$ & -23.91 & 3.20 & 3.9$^{+5.7}_{-2.3}$ \\
\hline
\end{tabular}
\end{center}
\end{table}

\begin{table}[t]
\centering
{\bf Table \ref{tab:sample} continued}
\begin{center}
\begin{tabular}{l c c c c c c}
\hline
Galaxy & Type & Core & M$_\mathrm{BH}$ & $K_\mathrm{s,sph}$ & $B-K_s$ & M$_\mathrm{sph,*}$ \\
& & & [$10^8$ M$_\odot$] & [mag] & [mag] & [$10^{10}$ M$_\odot$] \\
\hline
NGC4596 & SB0 & n & 0.79$^{+ 0.38}_{- 0.33}$ & -22.84 & 3.64 & 1.8$^{+2.7}_{-1.1}$ \\
NGC4621 & E & n & 3.9$^{+  0.4}_{-  0.4}$ & -24.49 & 3.66 & 8.4$^{+7.2}_{-3.9}$ \\
NGC4649 & E1 & y & 47$^{+   10}_{-   10}$ & -25.50 & 4.12 & 27$^{+23}_{-12}$ \\
NGC4697 & E4 & n &  1.8$^{+  0.2}_{-  0.1}$ & -24.06 & 3.78 & 6.0$^{+5.2}_{-2.8}$ \\
NGC4736 & Sab & n? & 0.060$^{+0.014}_{-0.014}$ & -21.54 & 3.17 & 0.43$^{+0.64}_{-0.26}$ \\
NGC4826 & Sab & n? & 0.016$^{+0.004}_{-0.004}$ & -22.47 & 3.11 & 1.0$^{+1.5}_{-0.6}$ \\
NGC4889 & cD & y & 210$^{+  160}_{-  160}$ & -27.00 & 4.41 & 122$^{+105}_{-57}$ \\
NGC4945$^{\dagger}$ & SBcd & n? & 0.014$^{+0.014}_{-0.007}$ & -20.60 & 4.21 & 0.30$^{+0.45}_{-0.18}$ \\
NGC5077 & E3 & y & 7.4$^{+  4.7}_{-  3.0}$ & -25.40 & 4.49 & 29$^{+25}_{-13}$ \\
NGC5576 & E3 & n & 1.6$^{+  0.3}_{-  0.4}$ & -24.50 & 4.24 & 11$^{+10}_{-5}$ \\
NGC5813 & E & y & 6.8$^{+  0.7}_{-  0.7}$ & -25.25 & 3.97 & 20$^{+17}_{-9}$ \\
NGC5845 & E3 & n & 2.6$^{+  0.4}_{-  1.5}$ & -22.99 & 4.25 & 2.8$^{+2.4}_{-1.3}$ \\
NGC5846 & E & y & 11$^{+    1}_{-    1}$ & -25.32 & 4.21 & 24$^{+20}_{-11}$ \\
NGC6086 & E & y & 37$^{+   18}_{-   11}$ & -26.50 & 4.42 & 78$^{+67}_{-36}$ \\
NGC6251 & E2 & y? & 5.9$^{+  2.0}_{-  2.0}$ & -26.57 & 4.73 & 96$^{+83}_{-44}$ \\
NGC6264 & S? & n? & 0.31$^{+0.004}_{-0.004}$ & -23.61 & 3.57 & 3.6$^{+5.2}_{-2.1}$ \\
NGC6323 & Sab & n? & 0.10$^{+0.001}_{-0.001}$ & -23.37 & 3.36 & 2.6$^{+3.8}_{-1.5}$ \\
NGC7052 & E & y & 3.7$^{+  2.6}_{-  1.5}$ & -25.97 & 4.73 & 55$^{+48}_{-26}$ \\
NGC7582 & SBab & n & 0.55$^{+ 0.26}_{- 0.19}$ & -22.83 & 3.06 & 1.4$^{+2.0}_{-0.8}$ \\
NGC7768$^{\dagger}$ & E & y & 13$^{+    5}_{-    4}$ & -26.41 & 4.21 & 64$^{+55}_{-30}$ \\
UGC3789 & SBab & n? & 0.11$^{+0.005}_{-0.005}$ & -22.66 & 3.18 & 1.2$^{+1.8}_{-0.7}$ \\
\hline
\end{tabular}
\end{center}
\begin{minipage}{3.25in}
$\dagger$: Galaxies for which we did not obtain new {\sc archangel} photometry. Column (1): Galaxy identifier. Column (2): Morphological type. Column (3): Core type. y indicates the galaxy contains a core, n indicates the galaxy does not have a depleted core. ? indicates that the classification is based on the velocity dispersion. Column (4): Supermassive black hole mass. References are provided in GS13. Column (5): Inclination and dust corrected {\it K$_s$} absolute spheroid magnitude. For elliptical galaxies the typical uncertainty on {\it K$_\mathrm{s,sph}$} is 0.25 mag. For disk galaxies this increases to 0.75 mag, due to the additional dust and bulge-to-total corrections. Column (6): ({\it B$-$K$_s$}) spheroid color. Column (7): Spheroid stellar mass. For elliptical galaxies the typical uncertainty on M$_\mathrm{sph,*}$ is 0.2 dex, for disk galaxies this increases to 0.36 dex.
\end{minipage}
\end{table}

In addition to the standard corrections mentioned above, two further `corrections' were applied to the absolute magnitude of disk galaxies to derive spheroid magnitudes. Given the large sample size, individual spheroid fractions were not derived for each object, but instead a mean statistical correction was applied based on each object's morphological type and disk inclination. The relationship between the applied correction (for both dust and bulge fraction) is based on the galaxy's morphological type and inclination and is given by equation (5) in GS13. The dust correction follows the method of \citet{Driver:2008} and depends on the galaxy inclination and passband. The correction for the bulge-to-disk flux ratio was derived from the observed bulge-to-disk flux ratios presented in \citet{Graham:2008c}, as adapted slightly by GS13. The bulge-to-disk correction depends on morphological type and passband, and is given in table 2 of GS13 for the {\it B-} and {\it K$_\mathrm{\it s}$}-bands.

While the above prescription results in significant uncertainties for the spheroid magnitudes of individual bulges, the ensemble average correction is considerably more accurate, scaling with $\sqrt{N}$. Such a statistical correction is only now viable given the sufficiently large sample of supermassive black hole masses in disk galaxies that are available in the literature. Our updated inclination and dust corrected {\it K$_\mathrm{\it s}$-}band spheroid magnitudes for the full sample are given in Table \ref{tab:sample}, the {\it B-}band values are given in table 2 of GS13.

We use the absolute galaxy magnitudes of the elliptical galaxies and the inclination and dust corrected bulge magnitudes of the disk galaxies to derive stellar masses for all spheroids in our final sample of 75 galaxies. Using the optical--NIR ({\it B$-$K$_s$}) color of the spheroids, we derive stellar mass-to-light ratios (M/L) using the relations presented in \citet{Bell:2001}. For the {\it K$_\mathrm{\it s}$-}band stellar mass-to-light ratio, M/L$_{K_s}$ and the ({\it B$-$K$_s$}) color, the relation is:

\begin{equation}
\log \mathrm{M/L}_{K_s} = -0.9586 + 0.2119 (\mathrm{\it B}-\mathrm{\it K}_s).
\end{equation}

We derive stellar mass-to-light ratios in the {\it K$_s$-}band as this ratio shows the smallest sensitivity to color in this band. For the range of {\it B$-$K$_s$} colors found in our sample, M/L$_{K_s}$ varies by a factor of 2, compared to a factor of 7 for the {\it B-}band mass-to-light ratio. {\it K$_s$-}band magnitudes also suffer the least from dust extinction, though as noted by \citet{Bell:2001} the conversion to stellar mass based on an optical-NIR color is not significantly affected by dust. The final stellar masses for all 75 spheroids in our sample are given in Table \ref{tab:sample}.

Lastly, we note that for all 75 galaxies, GS13 identified them as either S\'ersic or core-S\'ersic galaxies. For the majority of galaxies this classification was based upon examination of their surface brightness profiles taken from Hubble Space Telescope (HST) imaging. For 19 galaxies without HST imaging GS13 assigned a core type based on the galaxy's velocity dispersion (see GS13 for further details, including a list of those objects with velocity dispersion based core type assignments). The most massive galaxies are typically core-S\'ersic galaxies and the least massive galaxies are exclusively S\'ersic galaxies, however there is a significant region of overlap, with spheroids in the $10^{10}$--$10^{11} \rm{M}_\odot$ region showing both profile types. For all of these galaxies in the overlap region the classification was based on their observed light profile.

\subsection{Sources of uncertainty on M$_{\rm BH}$ and M$_{\rm sph,*}$}

The uncertainty on M$_{\rm BH}$ for each individual object is given in Table \ref{tab:sample}; these are drawn from the same sources as the black hole mass measurements themselves and have been adjusted to the distances tabulated in GS13 --- detailed references are given in GS13. The typical uncertainty on M$_{\rm BH}$ is 50\%. 

The total uncertainty on M$_{\rm sph,*}$ has contributions from two or three separate components. For elliptical galaxies these are the uncertainty on the {\it K$_s$} magnitudes measured by {\sc archangel} and the uncertainty in converting this magnitude into a stellar mass. The uncertainty in the magnitudes derived from the 2MASS photometry are determined by the {\sc archangel} pipeline, and are typically $\sim 0.25$ mags (or 0.1 dex). The uncertainty in the M/L used to convert these magnitudes to stellar masses depends on the uncertainty in the ({\it B$-$K$_s$}) color, and on the uncertain star formation history of each object. \citet{Bell:2001} give M/L$_{K_s}$ for a range of star formation histories, allowing the uncertainty in M/L due to the uncertain star formation history to be estimated. For our sample, the uncertainty on M/L$_{K_s}$ (due to both the uncertainty on the observed color and the uncertain star formation history) is typically 0.17 dex.  In elliptical galaxies the  typical total uncertainty on M$_{\rm sph,*}$ for disk-less systems is 0.2 dex.

In systems with a stellar disk there is the additional source of uncertainty in converting the total magnitude into a spheroid magnitude by applying both a dust correction and a correction for the bulge-to-disk flux ratio. The uncertainty due to the dust correction is estimated by \citet{Driver:2008} to be 5\% in both the {\it B-} and {\it K-}bands. The uncertainty in the bulge-to-disk flux ratio can be estimated from the data presented in \citet[their table 4]{Graham:2008c}, and for the galaxies in our sample is typically 0.3 dex. For systems with a disk this is the dominant source of uncertainty and the total typical uncertainty in M$_{\rm sph,*}$ is 0.36 dex.

\section{Analysis}
\label{sec:analysis}

\begin{figure}[t]
\includegraphics[width=3.5in]{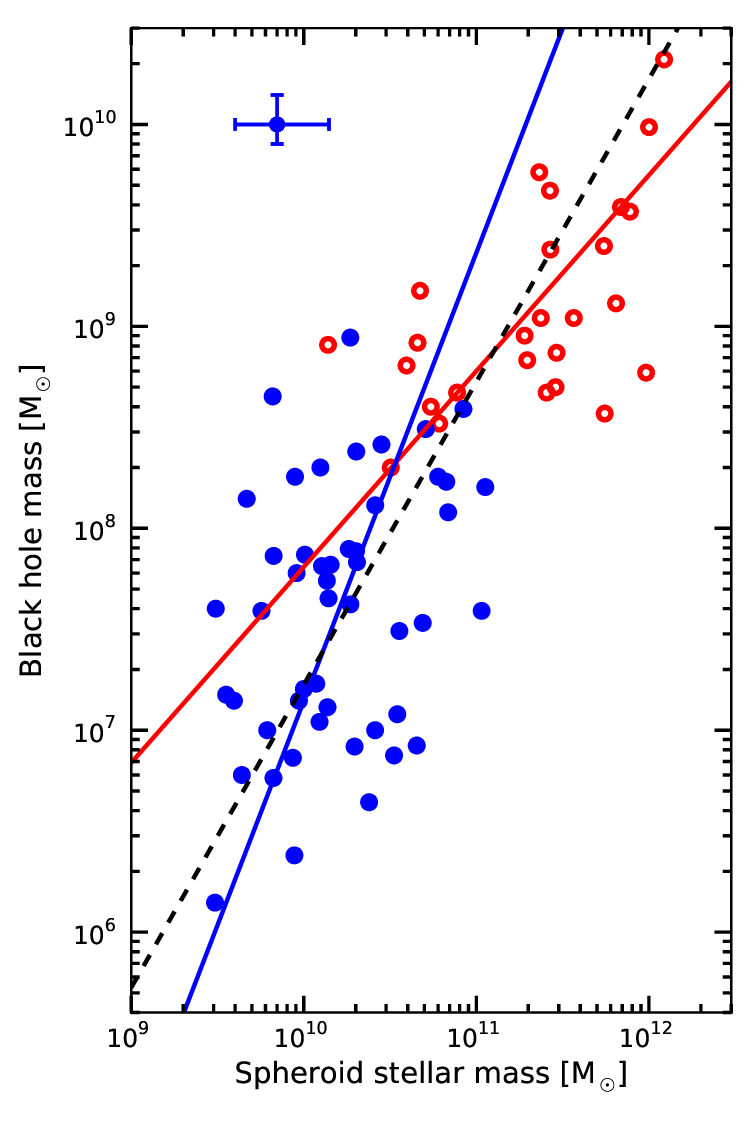}
\caption{Supermassive black hole mass vs. spheroid stellar mass for core-S\'ersic (open red symbols) and S\'ersic (filled blue symbols) galaxies. The best-fitting linear relations to the two samples are, given by Eqns. (\ref{eqn:core}) and (\ref{eqn:sersic}) shown as the solid lines. For comparison, the best-fitting linear regression for the full sample is shown as the dashed line and is dependent on the sample selection. A representative error bar is shown in the upper left corner.}
\label{fig:figure1}
\end{figure}

In Figure \ref{fig:figure1} we show the spheroid stellar mass plotted against the supermassive black hole mass for all galaxies in Table \ref{tab:sample}. We separate galaxies into S\'ersic (filled blue symbols) and core-S\'ersic (open red symbols). We fit separate linear regressions to the S\'ersic and core-S\'ersic galaxy subsamples, using the {\sc bces} {\it bisector} regression of \citet{Akritas:1996}. This technique takes into account the measurement uncertainties in both black hole mass and stellar spheroid mass and accounts for (though does not determine) the intrinsic scatter. For the core-S\'ersic galaxies the best-fitting symmetrical regression is:

\begin{align}
\label{eqn:core}
\log \frac{{\rm M}_{\rm BH}}{\rm{M}_\odot} &= (0.97 \pm 0.14) \log \left(\frac{{\rm M}_{\rm sph,*}}{3.0\times 10^{11}\ \rm{M}_\odot}\right) \nonumber \\
 &+ (9.27 \pm 0.09) ,
\end{align}
and for S\'ersic galaxies the best-fitting symmetrical regression is:
\begin{align}
\label{eqn:sersic}
\log \frac{{\rm M}_{\rm BH}}{\rm{M}_\odot} &= (2.22 \pm 0.58) \log \left( \frac{{\rm M}_{\rm sph,*}}{2.0\times 10^{10}\ \rm{M}_\odot}\right) \nonumber \\ &+ (7.89 \pm 0.18), 
\end{align}
with rms residuals of 0.47 and 0.90 dex respectively in the $\log {\rm M}_\mathrm{BH}$ direction. These linear relations are shown in Fig. \ref{fig:figure1} as the solid red and blue lines for the core-S\'ersic and S\'ersic galaxies respectively. For comparison, the linear regression to the combined sample is shown as the black dashed line, which has a slope of $1.50 \pm 0.12$ \citep[c.f.][]{Laor:2001}and an rms residual of 0.67 dex. The best-fitting regressions for the two different types of galaxy have significantly different slopes that are not consistent with each other given the confidence intervals on the slopes. S\'ersic galaxies follow an approximately quadratic relation, whereas core-S\'ersic galaxies follow an approximately linear relation. This is in agreement with the analysis and conclusions of \citet{Graham:2012a} who studied the M$_{\rm BH}$--M$_{\rm sph,dyn}$ relation.

The bend or break in the M$_\mathrm{BH}$--M$_\mathrm{sph,*}$ distribution occurs where the core-S\'ersic and S\'ersic relations overlap. This is at a spheroid stellar mass M$_\mathrm{sph,*} \sim 3\times10^{10}\ \rm{M}_\odot$, corresponding to a black hole mass M$_\mathrm{BH} \sim 2\times10^8\ \rm{M}_\odot$. The significance of this `break mass' will be discussed in Section \ref{sec:discussion}. Here we simply note that very few S\'ersic galaxies have M$_\mathrm{BH}$ greater than this break mass; equally no core-S\'ersic galaxy has a black hole mass below this value. 

\citet{Park:2012} examined the effect of using different linear regression techniques on the M$_\mathrm{BH}$--$\sigma$ relation but their results are applicable to supermassive black hole scaling relations in general. They reported that three popular regression techniques, {\sc bces} (as used in this work) a modified version of {\sc fitexy} \citep{Press:1992,Tremaine:2002} and a Bayesian technique developed by \citet{Kelly:2007}, {\sc linmix\_err} return consistent results. However, if the measurement uncertainties are larger than $\sim15$\% in the ordinate when using the `forward' regression, they find that the {\sc bces} routine may be biased to higher slopes \citep[as was noted by][]{Tremaine:2002}. Because of the significant uncertainties on many of our low-mass spheroid stellar masses (due to our statistical bulge-disk separation), we redetermined our core-S\'ersic and S\'ersic relations using the modified {\sc fitexy} and the {\sc linmix\_err} linear regression methods to test the robustness of our result. Following GS13, we determine a symmetrical bisector regression from the {\sc linmix\_err} routine by determining the `forward' and `inverse' regressions, then determining the line that bisects those two regressions. Using the {\sc linmix\_err} method we find bisector slopes for the core-S\'ersic and S\'ersic samples of $1.10 \pm 0.07$ and $2.11 \pm 0.46$ respectively.  We adopt a similar approach with {\sc fitexy}; determining the `forward' and `inverse' regressions then determining the bisector line. This method yields bisector regressions for the core-S\'ersic and S\'ersic samples with slopes of 1.08 and 2.48 respectively. With all three symmetric regression methods we find consistent best-fitting relations, and in all cases the slope for the S\'erisc galaxies is greater than $2 \sigma$ steeper than that for core-S\'erisc galaxies. We conclude that our principal result --- that core-S\'ersic and S\'ersic galaxies follow {\it different} M$_\mathrm{BH}$--M$_\mathrm{sph,*}$ relations is robust against the choice of linear regression method. 

The core-S\'ersic/S\'ersic classification is {\it similar} to the slow rotator/fast rotator (SR/FR) classification of \citet{Emsellem:2007}, though as discussed in \citet{Emsellem:2011} the overlap between the two systems is not perfect. Of our objects, 33/75 are part of the ATLAS$^\mathrm{3D}$ survey \citep{Cappellari:2011} and have FR/SR classifications from \citet{Emsellem:2011}. We also examined the spheroid stellar mass vs. supermassive black hole mass relation for these galaxies. We again fit linear regressions to the two samples of 9 SR and 24 FR galaxies. With this smaller sample we do not find significantly different slopes for the FR and SR samples; the slopes for the FR and SR samples are $1.71\pm0.27$ and $1.32\pm0.44$ respectively. While the slope for FRs is somewhat steeper, the difference is not significant given the formal uncertainty on the derived slopes. However, this may be caused by the galaxies with the most massive black holes and the least massive galaxies not having an FR/SR classification and therefore not being included in these regressions. This same sampling effect, where galaxies at the extremes of the supermassive black hole mass scaling relations were not well-sampled, led to a single log-linear relation being sufficient to describe the data in the past. It is only recently, where increased numbers of objects at the extremes of the relations have been added, has the bend in the supermassive black hole mass scaling relations become evident. Alternatively, the core-S\'ersic/S\'ersic classification may be more closely linked to the mechanism(s) responsible for black hole growth than the  FR/SR classification. Indeed, FR lenticular galaxies with depleted cores are known to exist \citep{Dullo:2012,Krajnovic:2013}. A larger sample of galaxies with both measured supermassive black hole masses and kinematic classifications would be desirable.

\begin{figure}
\includegraphics[width=3.5in,clip,trim= 50 10 30 10]{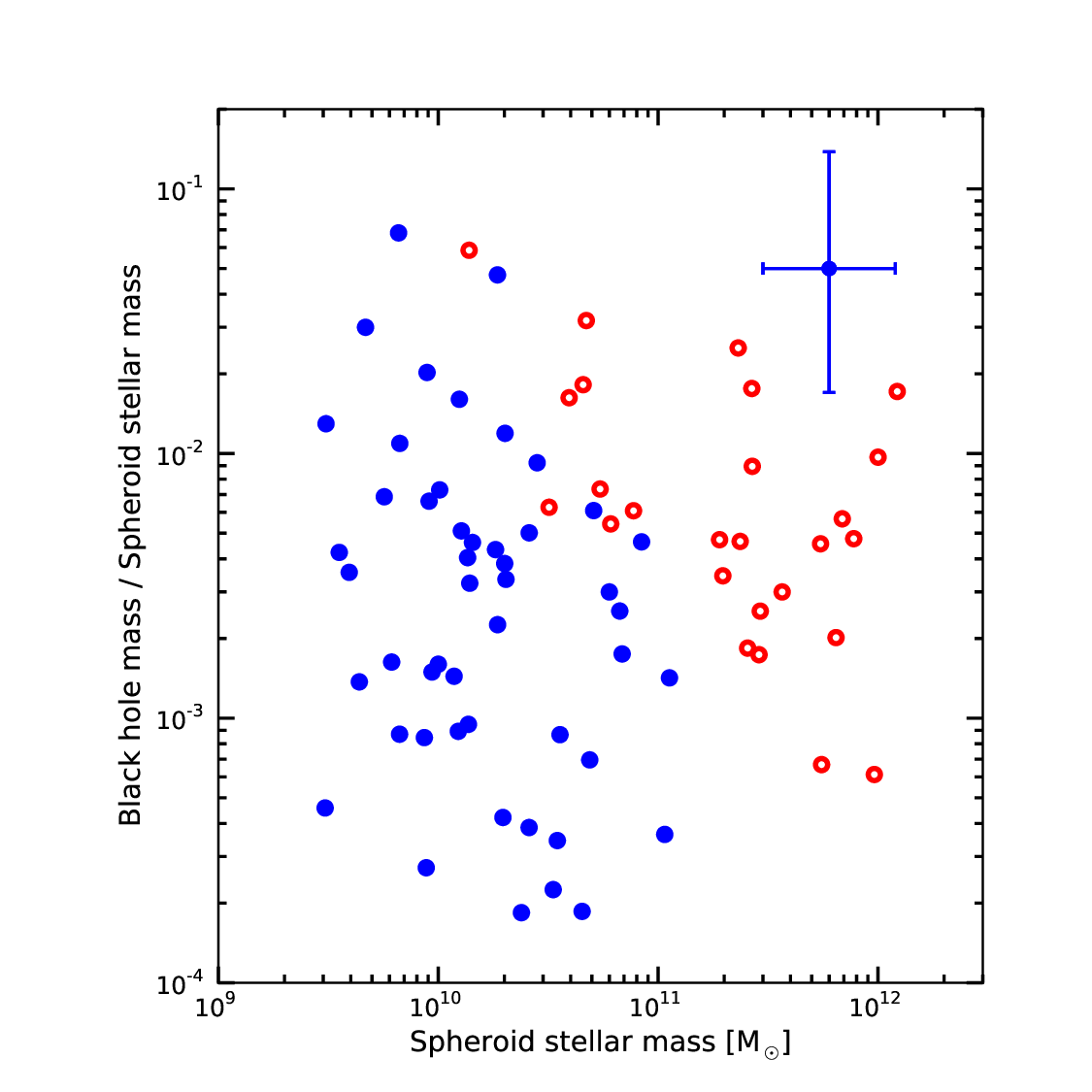}
\caption{The ratio of supermassive black hole mass to spheroid stellar mass, M$_\mathrm{BH}$/M$_\mathrm{sph,*}$ vs. M$_\mathrm{sph,*}$. As previously, core-S\'ersic galaxies are shown as red open symbols and S\'ersic galaxies as blue filled symbols. A representative error bar is shown in the upper right corner.}
\label{fig:figure3}
\end{figure}

From Eqn.\ (\ref{eqn:sersic}), the expected variation of M$_\mathrm{BH}$/M$_\mathrm{sph,*}$ with M$_\mathrm{sph,*}$ is close to linear for the S\'ersic galaxies, with the black holes of more massive spheroids representing a larger fraction of their host spheroid's mass. In Figure \ref{fig:figure3} we show the ratio of supermassive black hole mass to spheroid stellar mass as a function of spheroid stellar mass. In previous studies which found M$_\mathrm{BH} \propto {\rm M}_\mathrm{sph,*}^{\sim 1}$, this ratio was thought to be constant, with the supermassive black hole having a mass $\sim 0.15$--$0.20$\% of its host spheroid's mass \citep{Merritt:2001,Marconi:2003}. Here we find that this remains approximately true for the core-S\'ersic galaxies, albeit with M$_\mathrm{BH} \sim 0.55$\% of M$_\mathrm{sph,*}$ (the relation for core-S\'ersic galaxies is consistent with a slope of 1 hence a constant mass fraction). However, for S\'ersic galaxies, we find that the average M$_\mathrm{BH}$/M$_\mathrm{sph,*}$ is offset to a lower mean value of 0.3\% for our particular sample's mass range and it also displays a large range, from $\sim 0.02$--$2$\%. 

\section{Discussion}
\label{sec:discussion}
\subsection{Comparison to previous studies}

In this work we have identified a break in the M$_\mathrm{BH}$--M$_\mathrm{sph,*}$ diagram, caused by two separate log-linear relations with {\it significantly different slopes} for core-S\'ersic and S\'ersic galaxies. While this result is a significant change from the commonly accepted view of single log-linear scaling relations describing the relationship between black holes and their host spheroids, this work is not the first to identify this change. As noted in Section \ref{sec:intro}, \citet{Graham:2012a} and GS13 have both previously reported the need for separate log-linear relations  for core-S\'ersic and S\'ersic galaxies to describe the trend of black hole mass with host galaxy dynamical mass and host spheroid luminosity respectively. In addition, \citet{Graham:2007} had previously noted that both the M$_\mathrm{BH}$--$\sigma$ and M$_\mathrm{BH}$--L relations cannot both be log-linear due to the non-linear L--$\sigma$ relation for early-type galaxies.

While in this paper we have quantified the first bent relation between black hole mass and host spheroid stellar mass, we can compare our results with past work pertaining to the host's dynamical mass. GS13 reported an M$_{\rm BH}$--L$_{K_S}$ relation for the S\'ersic galaxies with a slope of $2.73\pm0.55$, which they equated to a slope of $2.34\pm0.47$ in the M$_{\rm BH}$--M$_{\rm dyn}$ diagram using M/L$_{K_s} \propto \mathrm{L}_{K_s}^{1/6}$ \citep[e.g.,][]{La-Barbera:2010,Magoulas:2012}. Prior to that, Graham (2012) had reported a slope of $1.92\pm0.38$ for the M$_{\rm BH}$--M$_{\rm dyn}$ relation based on independent data. These two steep slopes compare well with our slope of $2.22\pm0.58$ for the S\'ersic galaxy M$_{\rm BH}$--M$_{\rm sph,*}$ relation. For the core-S«ersic galaxies, GS13 had reported a slope of $1.10\pm0.20$ in the M$_{\rm BH}$--L$_{K_S}$ diagram. If core-S\'ersic galaxies are predominantly built by dry mergers of galaxies near or above the high-mass end of the S\'ersic distribution, then the slope in the M$_{\rm BH}$--M$_{\rm sph,*}$ diagram should be the same as the slope in the M$_{\rm BH}$--L diagram. Graham (2012) had additionally reported a slope of $1.01\pm0.52$ for the core-S\'ersic galaxies in the M$_{\rm BH}$--M$_{\rm dyn,sph}$ diagram (with the large uncertainty reflecting their small sample size).  These two shallower slopes are consistent with our slope of $0.97\pm0.14$ in the M$_{\rm BH}$--M$_{\rm sph,*}$ diagram.

Other authors have noted that low luminosity (or mass) galaxies are consistently offset from single log-linear black hole scaling relations derived from samples dominated by massive systems. \citet{Greene:2008} identified a population of low mass galaxies (stellar masses in the range $10^9$--$10^{10}\ \rm{M}_\odot$) whose black hole masses were offset below the M$_\mathrm{BH}$--M$_\mathrm{bulge}$ relation of \citet{Haring:2004} by an order of magnitude. More recently, \citet{Mathur:2012} determined black hole masses for a sample of 10 narrow-line Seyfert galaxies with host spheroid luminosities in the range $3\times10^9$--$3\times 10^{10}\ \rm{L}_\odot$ (corresponding to a range in stellar mass of approximately $10^{10}$--$10^{11}\ \rm{M}_\odot$) and again found their galaxies to be offset below the \citet{Haring:2004} relation. While \citet{Mathur:2012} attribute much of this offset to 5 of their objects being `pseudobulges', the remaining 5 are `classical' bulges and are still substantially offset from the M$_\mathrm{BH}$ - L$_\mathrm{bulge}$ relation of \citet{Gultekin:2009}. \citet{Greene:2008} also identify many of their objects which are significantly offset below the classical log-linear M$_\mathrm{BH}$--M$_\mathrm{bulge}$ relation as being well-fit with a de Vaucouleurs ($n=4$) profile, suggesting they are not pseudobulges. The offset objects found by both these studies are consistent with the log-linear scaling relations for S\'ersic galaxies reported in this work, \citet{Graham:2012a} and in GS13. While we do not identify pseudobulges in our sample, we note that both NGC4486a and NGC821, both `classical' elliptical galaxies with no indication of a pseudobulge, are consistent with our S\'ersic scaling relation and offset from the core-S\'ersic relation, arguing against the offset nature of the low-mass systems being a pseudobulge phenomena.

\citet{Sani:2011}, \citet{Vika:2012} and \citet{Beifiori:2012} have all recently constructed M$_\mathrm{BH}$--M$_\mathrm{sph,*}$ relations \citep[or in the case of ][the closely-related M$_\mathrm{BH}$--L$_{K,\mathrm{sph}}$ relation]{Vika:2012} for large samples ($\sim 50$ galaxies). All three studies only considered single log-linear fits to their data and are dominated by massive galaxies with M$_\mathrm{sph,*} > 3\times10^{10}\ \mathrm{M}_\odot$. As expected from their high-mass-dominated samples, all three studies find M$_\mathrm{BH}$--M$_\mathrm{sph,*}$ relations with slopes $\sim1$, consistent with our finding for core-S\'ersic galaxies. However, in all three studies a number of galaxies are offset below those authors' single log-linear relations. With host spheroid masses around $3\times10^{10}\ \rm{M}_\odot$, they are consistent with the high-mass regime of our S\'ersic galaxy relation. These studies highlight the need to extend the range of supermassive black hole masses and host spheroid masses used to examine the black hole scaling relations.

\subsection{Scatter about the M$_\mathrm{BH}$--M$_\mathrm{sph,*}$ relation}
As well as determining the slope of the supermassive black hole scaling relations, many studies also examine the scatter about the relations. In particular, this is done to argue that one of the relations has reduced intrinsic scatter compared to the other common scaling relations, and therefore is the `fundamental' driving relation. At low masses the scatter in our Figure \ref{fig:figure1} is dominated by the large measurement uncertainties due to our statistical dust and bulge correction, making such a comparison difficult for this study. Instead, we will make a few simple observations on the revised expectations for the {\it intrinsic} scatter as a result of our bent scaling relation.

As noted in Section \ref{sec:analysis}, the observed scatter (the sum of the intrinsic scatter and measurement uncertainty) in the black hole mass direction is significantly larger for the S\'ersic galaxies than for the core-S\'ersic galaxies: 0.90 dex compared to 0.47 dex. An increase is expected given the typically larger measurement uncertainties for the S\'ersic galaxies due to the statistical dust and bulge correction. However, we also expect the scatter in the vertical direction to increase because of the increased slope of the relation for the S\'ersic galaxies. This is not the case for the intrinsic scatter orthogonal to the relation --- we would expect this to be reduced relative to a single log-linear relation fit to the entire data. The orthogonal scatter we find for the core-S\'ersic galaxies is 0.26 dex, and for the S\'erisc galaxies it is 0.29 dex, an improvement over the 0.31 dex orthogonal scatter we find for the single log-linear relation.

Finally, a number of theoretical studies have examined the idea of the supermassive black hole scaling relations being the product of the repeated merging of a randomly seeded initial population of black holes and host galaxies \citep{Peng:2007,Jahnke:2011}. These studies predict a decrease in the intrinsic scatter as host galaxy mass increases \citep[though][note that the addition of star formation to their model reduces the rate of this decrease in scatter]{Jahnke:2011}. Given the large measurement uncertainties for many of our S\'ersic galaxies it is difficult to quantify any variation of the intrinsic scatter with host mass in our sample. Moreover, we expect that only for the core-S\'ersic galaxies is the binary merging of both galaxies and black holes the dominant mechanism of growth. As these galaxies only span a relatively narrow range in host spheroid mass, any systematic variation of the intrinsic scatter within this subsample is uncertain.

\subsection{Comparing supermassive black hole and nuclear star cluster scaling relations}

\begin{figure}
\includegraphics[width=3.25in,clip,trim= 50 10 30 10]{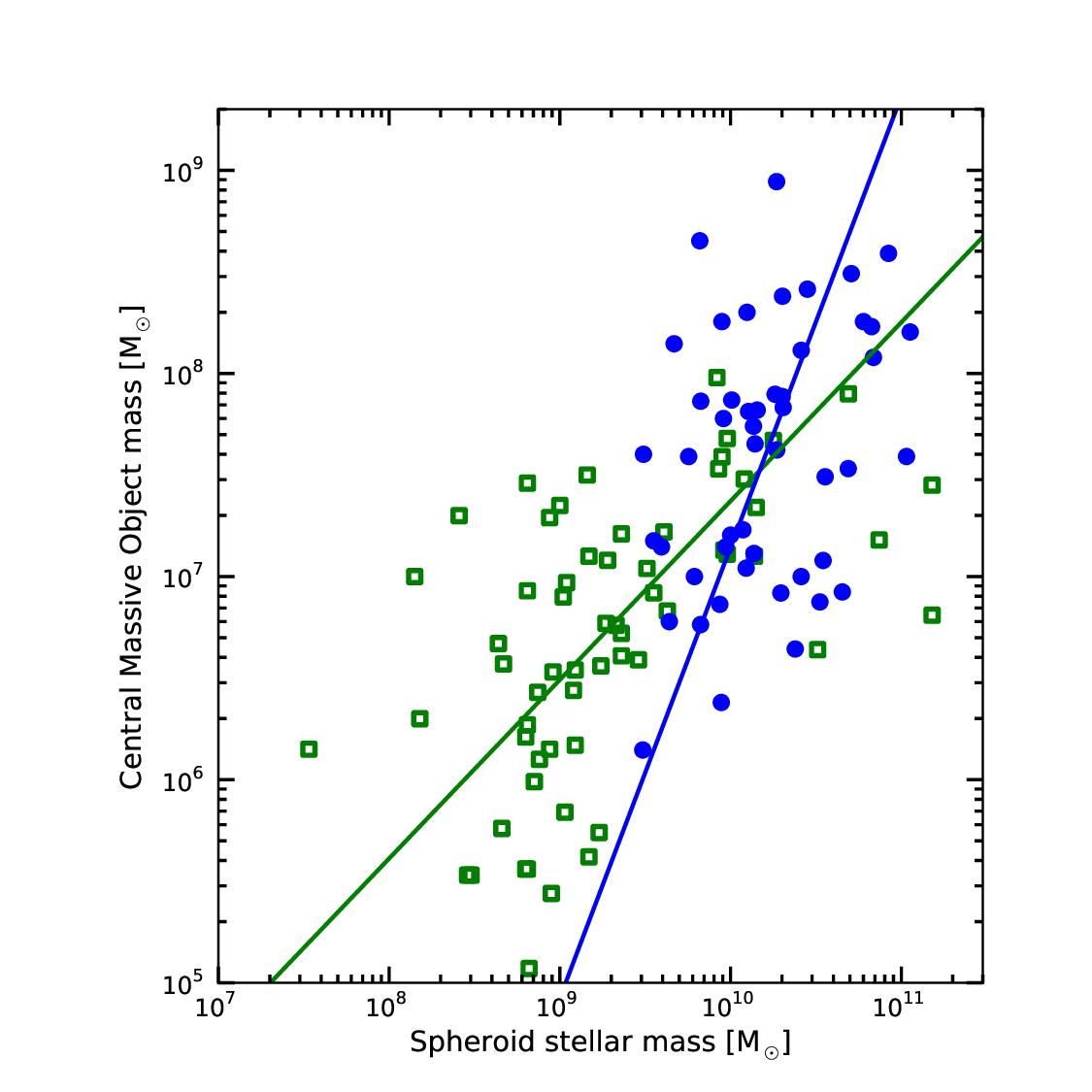}
\caption{Central massive object mass, M$_\mathrm{CMO}$ vs. spheroid stellar mass, M$_\mathrm{sph,*}$ for our S\'ersic galaxies (blue points) and for the nuclear star clusters (and host spheroids) of \citet[open green squares]{Scott:2012}. The blue and green lines are the best-fitting symmetrical linear relations to the two samples (the green line is taken from \citet{Scott:2012}).}
\label{fig:figure4}
\end{figure}

\citet{Ferrarese:2006a} and \citet{Wehner:2006} have argued that nuclear star clusters and supermassive black holes form a single class of central massive object (CMO) based on their allegedly common mass scaling relations, though recent studies \citep{Balcells:2007,Graham:2009,Graham:2012b,Erwin:2012,Leigh:2012,Scott:2012} have since argued against this scenario. We briefly re-examine the connection between nuclear star cluster and supermassive black hole scaling relations in the light of the bent M$_\mathrm{BH}$--M$_{\rm sph,*}$ relation. In Figure \ref{fig:figure4} we show M$_\mathrm{CMO}$ vs. M$_\mathrm{sph,*}$ for the S\'ersic galaxies presented in this study and the nuclear star clusters presented in \citet{Scott:2012}. The two lines show the best-fitting linear relations to the two samples. The M$_{\rm NC}$--M$_{\rm sph,*}$ line is taken from \citet{Scott:2012} and has a slope of $0.88 \pm 0.19$. The relation for the nuclear star clusters is $2.3 \sigma$ shallower than that for the supermassive black holes. This difference is more pronounced than that reported by \citet{Scott:2012}, due to our improved bent supermassive black hole scaling relation, strongly suggesting that supermassive black holes and nuclear star clusters do not follow a common scaling relation and therefore do not have a common formation mechanism.

\subsection{The growth of supermassive black holes and stellar spheroids}

As noted in Section \ref{sec:analysis}, the break or bend in the M$_\mathrm{BH}$--M$_\mathrm{sph,*}$ relation occurs at M$_\mathrm{sph,*} \simeq 3\times10^{10}\ \rm{M}_\odot$. This mass scale has been identified by a number of authors as marking a change in early-type galaxy properties. Below this mass galaxies are typically young, low-surface density objects \citep{Kauffmann:2003}, whose spheroid's surface brightness is well-fit by a S\'ersic profile \citep{Graham:2003,Balcells:2003}, and define a sequence of increasing M/L and bulge fraction \citep{Cappellari:2012} --- until dwarf spheroidal systems appear around $M_B \sim -14$ mag. Above this mass galaxies are typically old with high S\'ersic indices \citep{Graham:2001,Kauffmann:2003}, have core-S\'ersic surface brightness profiles \citep{Graham:2003}, are bulge-dominated and show only a narrow range in M/L \citep{Cappellari:2012}. \citet{Tremonti:2004} also report that the mass-metallicity relation flattens above this mass scale.

The results discussed above are all consistent with a scenario in which galaxies with spheroid masses below the rough $3\times10^{10}\ \rm{M}_\odot$ divide grow predominantly through gas-rich processes involving significant in-situ star formation whereas the growth of those galaxies above this mass scale is dominated by `dry' processes involving little gas and the direct accretion of stars via mergers. This is similar to a scenario adopted by \citet{Shankar:2012} based on their semi-analytic model of black hole and galaxy formation. 

In the high-mass regime where little gas is involved, black hole growth will be dominated by the binary merger of supermassive black holes during galaxy mergers. In a dry merger, to first order, the masses of the two progenitor galaxies will simply be added, as will the masses of their supermassive black holes. Thus, in this regime we can roughly expect a linear relationship between supermassive black hole mass and host stellar mass, as observed.

In the low mass regime gas plays a more significant role, complicating the picture. Galaxies can grow from the direct cooling of gas from their immediate surroundings, and even when mergers do play a significant role the resulting growth (of both stellar mass and black hole mass) will not be simply additive as the `wet' mergers will likely be accompanied by significant star formation. The slope of $\sim 2.2$ in the M$_\mathrm{BH}$--M$_\mathrm{sph,*}$ relation that we find for S\'ersic galaxies implies that black holes have grown more rapidly than their host spheroids. This is consistent with the growth of these galaxies being dominated by gaseous processes \citep[e.g., gas-rich mergers, cooling of hot gas, secular evolution: see ][for a detailed discussion of the full range of processes]{Hopkins:2009} that efficiently channel material onto the central black hole \citep[see e.g.,][]{Marconi:2004} but are less effective at growing the host galaxy spheroid. One such example of a supermassive black hole growing more rapidly than its host galaxy has recently been observed by \citet{Seymour:2012}. This steepening of the supermassive black hole mass -- host spheroid stellar mass relation has been reproduced in a number of simulations \citep[][figures 5, 3 and 8 respectively]{Cirasuolo:2005, Dubois:2012, Neistein:2013}, though these authors have typically focused on the agreement with the classical single log-linear relation in massive galaxies, rather than the downturn at lower mass  in their simulated data. 

\section{Conclusions}
\label{sec:conclusions}
The results in Section \ref{sec:analysis} represent a major revision to the accepted picture of the relationship between supermassive black holes and their host galaxies. Our results have significant implications for:  studies of how supermassive black holes and their hosts co-evolve over cosmic time; the detection of intermediate-mass supermassive black holes; and models of feedback from active galactic nuclei due to supermassive black hole growth.

We have examined the bent nature of the supermassive black hole mass -- spheroid stellar mass relation for a large sample of 75 galaxies. We find that, when separating galaxies into core-S\'ersic and S\'ersic galaxies based on their central light profiles, they follow significantly different supermassive black hole scaling relations whose slopes are {\it not} consistent with one another. For the core-S\'ersic galaxies M$_\mathrm{BH} \propto {\rm M}_\mathrm{sph,*}^{0.97\pm0.14}$, whereas for the S\'ersic galaxies M$_\mathrm{BH} \propto {\rm M}_\mathrm{sph,*}^{2.22\pm0.58}$. These results are consistent with the expectation from a single log-linear supermassive black hole mass -- host velocity dispersion relation combined with a bent relationship between host spheroid luminosity and host velocity dispersion. We find that the supermassive black hole mass is an approximately constant 0.55\% of a core-S\'ersic galaxy's spheroid stellar mass. The non-log-linear M$_\mathrm{BH}$--M$_\mathrm{sph,*}$ relation implies that, for S\'ersic galaxies, M$_\mathrm{BH}$ {\it is not} a constant fraction of the host spheroid mass, but represents an increasing fraction with increasing M$_\mathrm{sph,*}$. 

\acknowledgements
This research was supported by Australian Research Council funding through grants DP110103509 and FT110100263. 

\bibliographystyle{apj}
\bibliography{black_holes}

\end{document}